\documentclass[twocolumn]{openjournal}

\usepackage[utf8]{inputenc}
\usepackage[T1]{fontenc}
\usepackage{float}
\usepackage{xcolor}
\usepackage{ulem}
\usepackage{graphicx}	
\usepackage{amsmath}	
\usepackage{comment}
\usepackage{color}
\usepackage{hyperref}
\hypersetup{colorlinks=true,
	urlcolor=blue,  
	linkcolor=blue,  
	citecolor=blue, 
    menucolor=blue, 
    urlcolor=blue}  

\def\prd{{\rm Phys. Rev. D}}
\def\apj{{\rm ApJ}}
\def\aj{{\rm AJ}}
\def\apjs{{\rm ApJS}}

\def\mnras{{\rm MNRAS}}
\def\sqdeg{\,{\rm deg ^2}}
\def\persqdeg{\,{\rm deg ^{-2}}}

\def\angs{\textrm{ \AA}}

\def\hmpc{h^{-1}{\rm Mpc}}
\def\hgpc{\;h^{-1}{\rm Gpc}}
\def\invhmpc{\;h\;{\rm Mpc}^{-1}}

\def\lya{Ly$\alpha$ }

\def\simlt{\lower.5ex\hbox{$\; \buildrel < \over \sim \;$}}
\def\simgt{\lower.5ex\hbox{$\; \buildrel > \over \sim \;$}}

\begin{document}

\title{Forecasting the Detection of Lyman-alpha Forest Weak Lensing from the Dark Energy Spectroscopic Instrument and Other Future Surveys}

\author{Patrick Shaw$^{1,2,*}$}
\author{Rupert A.C. Croft$^{1,2}$}
\author{R. Benton Metcalf$^{3,4}$}
\thanks{$^*$E-mail:pshaw2@andrew.cmu.edu}
\affiliation{$^{1}$ McWilliams Center for Cosmology, Dept. of Physics, 
Carnegie   Mellon  University, Pittsburgh, PA 15213, USA
\\
$^{2}$ NSF AI Planning Institute for Physics of the Future, 
Carnegie   Mellon  University, Pittsburgh, PA 15213, USA
\\
 $^{3}$ Dipartimento di Fisica \& Astronomia, Universit\'{a} di Bologna, via Gobetti 93/2, 40129 Bologna, Italy
\\
$^{4}$ INAF-Osservatorio Astronomico di Bologna, via Ranzani 1, 40127 Bologna, Italy
\\
}

\begin{abstract}
The apparent angular positions of quasars are deflected on the sky by the gravitational field sourced by foreground matter. This weak lensing effect is measurable through the distortions it introduces in the lensed quasar spectra. Discrepancies in the statistics of the Lyman-$\alpha$ forest spectral absorption features can be used to reconstruct the foreground lensing potential. We extend the study of this method of Lyman-$\alpha$ forest weak gravitational lensing to lower angular forest spectrum source densities than previous work. We evaluate the performance of the Lyman-$\alpha$ lensing estimator of \cite{ben1} on mock data based on the angular forest source density ($50\persqdeg$) and volume ($\sim$700,000 spectra total) of the DESI survey. We simulate the foreground galaxy distribution and lensing potentials with redshift evolution approximated by N-body simulation and simulate Gaussian-random Lyman-$\alpha$ forests to produce mock data for the entire DESI footprint. By correlating the foreground galaxy distribution with the potential reconstructed by the estimator, we find that a weak lensing detection with signal to noise of $\sim4$ will be possible with the full DESI data. We show that spectral surveys with low density and high volume are promising candidates for forest weak lensing in addition to the high resolution data that have been considered in previous work. We present forecasts for future spectral surveys and show that with larger datasets a detection with signal to noise $>10$ will be possible.
\end{abstract}


\maketitle



\section{Introduction}
\label{intro}
Weak gravitational lensing has emerged as one of the most efficacious methods of probing the matter density of the Universe. Its direct sensitivity to both baryonic and dark matter make it an invaluable tool in constraining models of cosmology (\citealt{lensingreview,lensing_review_3,lensing_review_2}). 
Gravitational lensing broadly refers to the diverse set of methods used to measure the deflections in the paths of light rays emitted from distant astrophysical sources. These deflections, caused by the gravitational effects of intervening mass, encode information about the matter distribution along the line of sight. In the weak lensing regime, the lensing distortions are sufficiently small that they can only be detected through statistical methods applied to large data sets. Most commonly, weak lensing is studied in the context of large galaxy surveys wherein distortions in the shapes of galaxy images can be used to reconstruct the foreground lensing mass distribution (\citealt{galaxy_lensing,galaxygalaxy_lensing2}). However, methods have also been developed for the cases of continuous fields such as the CMB (\citealt{bernard97,1997ApJ...489....1M,1998ApJ...492L...1M,zald99,huok02,cmbestimator,schaan19}) and 21 cm line radiation (\citealt{2007MNRAS.381..447M,2009MNRAS.394..704M,21cmlensing}). In these cases, lensing is measured by detecting deviations in the statistics of the field. In this work, we will discuss how weak lensing can be measured in the Lyman-alpha forest (\lya forest), a novel continuous field.

The \lya forest is observed in the spectra of quasars and high redshift ($z\sim 3$) galaxies (\citealt{rauch98,prochaska19}). It is a set of absorption features which encode the relative density of neutral hydrogen in the intergalactic medium (IGM) along the line of sight to the source of the spectrum. Neutral hydrogen absorbs incident radiation with high efficiency at the wavelength corresponding to the \lya transition in the hydrogen atom. As the light from the source is redshifted, each wavelength observed corresponds to a particular redshift where the \lya absorption occurs. Therefore, the magnitude of the absorption feature maps the neutral hydrogen density at that redshift, ultimately yielding a one-dimensional sampling of the hydrogen density along the line of sight. In the \lya forest context, one typically works with the flux overdensity or flux transmission field (\citealt{lya_tomography_deltaflux,lya_powerspectrum_deltaflux}), $\delta_f\left(\lambda\right)$, which is defined as
\begin{align}
\label{deltafluxdefinition}
\delta_F\left(\lambda\right) \equiv \frac{F\left(\lambda\right)}{\left< F \right>\left(\lambda\right)} -1.
\end{align}
The measured flux, $F$, at a wavelength $\lambda$, is divided by the expected flux $\left< F \right>$, where the expected flux is the radiation at that given wavelength due to the unabsorbed quasar or galaxy spectrum as well as the background. The result is a quantity with mean zero that is a biased tracer of the hydrogen density contrast along the line of sight.
When many \lya sources are combined from a large spectral survey, the \lya forest features measured can be combined to recover a sampling of the three-dimensional continuous flux overdensity field. 

This field is a promising candidate for weak lensing because it is well-understood, easily-simulated, and large observational data sets are already available with more on the way (\citealt{eBOSS,DESI,CLAMATO,MSE,LATIS,subaruPFS}). Additionally, redshift measurements are very accurate, leading to negligible errors in the lensing geometry; it has different sources of systematic error compared to galaxy lensing as it does not rely on ellipticity models or suffer from the instrumental and atmospheric effects that impact galaxy lensing studies (\citealt{lensing_review_2}); and it is more information-dense due to its three-dimensional nature. While CMB lensing also has the advantage of precise redshift measurement, it is only two-dimensional and probes a higher redshift, $z \sim 3$ (\citealt{ben2,CMBkernel}), compared to \lya lensing which is most sensitive to $z\sim 1$. Therefore, \lya lensing has the potential to be a source of high quality measurements of structure at a unique epoch of the Universe, complementing existing methods. 

However, the \lya forest introduces some challenges compared to other lensing sources due to its irregular geometry. While the flux transmission field is three-dimensional and continuous, its measurement is limited by the positions of backlight sources. This means that it is not sampled regularly. This prohibits the use of the Fourier-space based methods used for other continuous fields, as these methods depend upon uniform sampling to be effective. In \cite{ben1}, an estimator was derived that is resilient to the sparse geometries of the \lya forest field. We are also limited by the relatively high amount of noise in \lya flux measurements. Surveys that are large enough to measure lensing typically have individual pixel signal to noise well below unity (\citealt{eBOSS,CLAMATO,LATIS,DESI_EDR}). However, we are hopeful that the large number of spectra in DESI will be able to overcome this difficulty.

Previous work (\citealt{ben1,patrick}) has produced forecasts for weak lensing detection from \lya forest data by simulating small (field area  $\sim 1\sqdeg$ or smaller), high source density (angular \lya source density typically  $\sim1000\persqdeg$) using simple Gaussian random field models for the lensing potential. In this work, our goal is to extend these methods to more realistic potentials based on a forthcoming observational dataset which has a lower \lya source density over a much larger footprint (50 $\persqdeg$ over 14,000 $\sqdeg$).

In this work, we apply the aforementioned \lya forest lensing estimator to data sets simulating the Dark Energy Spectroscopic Instrument (DESI)  survey, a large spectroscopic survey which already has preliminary data publicly available in the form of the Early Data Release (EDR) (\citealt{DESI_EDR}), with full data being available in 2026 (\citealt{DESI_schedule}). We make predictions about the expected signal to noise (S/N) of a weak lensing detection from DESI as well as providing a forecast for the S/N that would be achievable with future surveys. Due to the low angular density of \lya sources in DESI, direct reconstruction of a single lensing potential field with high S/N is not possible, so we instead detect weak lensing by correlating our estimator reconstructions with mock foreground galaxy distributions. An outline of our approach is as follows: first we simulate a DESI survey volume, from which we compute an exact lensing potential as well as a mock galaxy survey. We then simulate \lya forests consistent with the geometry of DESI and lens those according to the exact lensing potential obtained from the foreground simulation. Our lensing estimator produces reconstructions of the lensing potential from the lensed forest data, which is then correlated with lensing potentials that are obtained from the mock galaxy surveys (data which will also be available from DESI). With this procedure we predict the S/N of a weak lensing detection from the survey as a whole. 

An outline of our paper is as follows: We will first provide a brief overview of the weak lensing formalism and observables, and then a description of our quadratic estimator. We will then explain how we simulate mock data for both the DESI foreground and \lya forest, as well as how we apply the quadratic estimator to these data. Next, we present our results including a prediction for the S/N of weak lensing detection from DESI, discussion of how we address bias in the estimates, and predictions for the improvement we expect to find with larger estimator redshift bins. Finally, we will provide a forecast for our expectations regarding lensing measurements from forthcoming surveys and discuss our overall predictions for the future of \lya lensing. 

\section{Estimating weak Lensing in the \lya Forest}
\label{method}
\subsection{Lensing formalism}
\label{lensing_formalism}
The primary observable in this work is the lensing potential,
\begin{align}
    \label{potfromkappa_equation}
    \phi(\boldsymbol{\theta})=\frac{1}{\pi} \int_{\mathbb{R}^{2}} \mathrm{~d}^{2} \theta^{\prime} \kappa\left(\boldsymbol{\theta^{\prime}}\right) \ln \left|\boldsymbol{\theta}-\boldsymbol{\theta^{\prime}}\right|,
\end{align}
which we define in terms of the dimensionless convergence, $\kappa( \boldsymbol{\theta})$, in angular coordinates. The convergence can be calculated through an integral of the matter density contrast, 
\begin{align}
\label{delta_rho_equation}
\delta_{\rho}( \boldsymbol{\theta},\chi) = \frac{\rho( \boldsymbol{\theta},\chi)}{\left<\rho( \boldsymbol{\theta},\chi)\right>} -1,
\end{align}
multiplied by the lensing kernel, over the comoving distance to the forest, $\chi_s$.
\begin{align}
\label{kappa_equation}
    \kappa(\boldsymbol{\theta})=\frac{3}{2} \frac{\Omega_{m} H_{0}^2}{c^{2}} \int_{0}^{\chi_{s}} d \chi\left[\frac{d_{A}(\chi) d_{A}\left(\chi, \chi_{s}\right)}{d_{A}\left(\chi_{s}\right)a}\right] \delta_\rho(\boldsymbol{\theta}, \chi),
\end{align}
where $d_{A}(\chi)$ is comoving angular size distance, $H_0$ is the Hubble constant, $\Omega_m$ is the matter density parameter, and $a$ is the scale factor. The convergence is related to the lensing potential by a Poisson equation, 
\begin{align}
    &\nabla^{2} \phi(\boldsymbol{\theta})=2 \kappa(\boldsymbol{\theta}),
\end{align}
and the lensing potential is related to the angular deflection field, $\boldsymbol{\alpha}( \boldsymbol{\theta})$, by a gradient
\begin{align}
    &\boldsymbol{\alpha}(\boldsymbol{\theta})=\nabla\phi(\boldsymbol{\theta}),
\end{align}
hence why it is thought of as a potential. The above analysis makes the assumptions of both the thin lens and Born approximations (see  \cite{lensingreview} for more details). The lensing potential is of great scientific interest because it is one of the few direct measurements of the total matter distribution. Because it is related explicitly to $\delta_\rho(\boldsymbol{\theta})$, which is predominantly dark matter, it probes large scale structure with few assumptions and is therefore very valuable for constraining cosmological models.

\subsection{Quadratic estimator}
\label{estimatortheory_section}
Broadly speaking, weak lensing of continuous fields is observed by measuring deviations from the expected statistics of the field in question, specifically the correlation function. For this work we assume the model for the \lya forest flux field power spectrum proposed in \cite{McDonald}, and use it to compute the forest's expected correlation function at $z=2.5$. It is important to note that any error in the assumed correlation function would bias the estimator proposed. However, on relevant scales ($k< 10 \invhmpc$), errors in the \lya correlation function model of \cite{McDonald} are below 5\% (\citealt{lyamodel_test}) according to tests using hydrodynamic simulations. The Lyman-alpha forest flux overdensity field is a biased tracer of the underlying matter density and typically has smaller errors than the matter power spectrum. The forest bias parameter increases as $k$ decreases, so this is especially true at the large scales we are considering. The errors in the assumed power spectrum are well below the S/N of our study. Therefore, we conclude that for our purposes this model is sufficient.

The quadratic estimator we use (derived in much more detail in \citealt{ben1}) estimates the lensing potential (Eq. \ref{potfromkappa_equation}) in the basis of Legendre polynomials,
\begin{align}
\label{legendre expansion}
    \phi(\boldsymbol{\theta})=\sum_{m=0}^{N_{x}} \sum_{n=0}^{N_{y}} \hat{\phi}_{mn} P_{m}(x) P_{n}(y).
\end{align}
The $P_n$ are the Legendre polynomials, and the variables are scaled such that
\begin{align}
x \equiv 2 \left(\frac{\theta_{1}-\theta_{1}^{o}}{\Delta \theta_{x}}\right)-1, \quad y \equiv 2 \left(\frac{\theta_{2}-\theta_{2}^{o}}{\Delta \theta_{y}}\right)-1,
\end{align}
where the $\left(\theta_1,\theta_2\right)$ are the angular coordinates of the field origin (lower left) and the $\Delta \theta_{x,y}$ are the field widths. The basis of Legendre polynomials is chosen due to its advantages over Fourier-space based methods for the case of the \lya forest. It allows us to truncate the estimate at an arbitrary scale without imposing periodic boundary conditions. For more details about the derivation of our estimator, see \cite{ben1}. Briefly, the estimates for the parameters $\hat{\phi}_\mu$ (where $\mu$ indexes over $m$ and $n$) are given by
\begin{align}
\label{estimator_eq}
    \hat{\phi}_{\mu}=\frac{1}{2} F_{\mu \nu}^{-1}\left(\boldsymbol{\delta}^{\top} \mathbf{C}^{-1} \mathbf{P}^{* \nu} \mathbf{C}^{-1} \boldsymbol{\delta}-\rm{tr}\left[\mathbf{C}^{-1} \mathbf{P}^{* \nu}\right]\right),   
\end{align}
where $F_{\mu\nu}^{-1}$ is the inverted Fisher matrix, $\boldsymbol{\delta}$ is a vector of the \lya forest flux overdensities (as defined in Eq. \ref{deltafluxdefinition}), $\mathbf{C}$ is the covariance matrix between \lya flux overdensity pixels including intrinsic correlations and noise (assumed based on the correlation function derived from the model of \citealt{McDonald}, which relates the underlying matter density to the biased \lya forest flux field), and $\mathbf{P}$ is constructed from the derivatives of the Legendre polynomials. For details about the construction of $\mathbf{P}$ in various bases, please consult the appendices of \cite{ben1}. In this work we use the Legendre polynomial basis. It can be shown that the covariance of this estimator is consistent with the Cramér-Rao lower limit, implying it is efficient. This estimator is discrete rather than continuous, meaning it can accommodate the irregular geometries associated with the one-dimensional sampling of \lya forest data described in Section \ref{intro}. In practice, the estimator that we calculate is a block diagonal version of Eq. \ref{estimator_eq}, where we consider only correlations between pixels in the same redshift bin. The correlation function in this case is
\begin{align}
\boldsymbol{C}=\left(\begin{array}{cccc}\boldsymbol{C}_{(1)} & 0 & \cdots & 0 \\ 0 & \boldsymbol{C}_{(2)} & \cdots & 0 \\ 0 & 0 & \ddots & 0 \\ 0 & 0 & \cdots & \boldsymbol{C}_{(n)}\end{array}\right),
\end{align}
where the subscripts denote the different redshift bins. The estimator then becomes
\begin{align}
\label{binnedestimator_equation}
\hat{\phi}_\mu&=\frac{1}{2} F_{\mu v}^{-1} \sum_{k=1}^n \operatorname{tr}\left[\mathbf{C}_{(k)}^{-1} \mathbf{P}_{(k)}^v\left(\mathbf{C}_{(k)}^{-1} \boldsymbol{\Delta}_{(k)}-\mathbf{I}\right)\right]\\
&=\frac{1}{2} F_{\mu \nu}^{-1} \sum_{k=1}^n \tilde{\phi}_v^k,
\end{align}
where $n$ is the number of redshift bins. With this method, we ignore correlations between pixels in different redshift bins under the assumption that forest correlations grow weaker at longer distances and not much useful information is being lost. In \cite{ben1}, this assumption is shown to be true empirically for the geometries adopted in that work. This approach is necessary because it reduces the memory cost and compute time of calculating the estimator by a factor of $\sim1/n^2$. In section \ref{longersightlines_section}, we discuss the possibility of relaxing this assumption to recover more information. For the case of uniform noise, Eq. \ref{binnedestimator_equation} reduces to a simple average of the estimates from each redshift bin. Each wavelength bin in the quasar spectra corresponds to a particular redshift and therefore a plane of \lya pixels. We refer to the size of the redshift bins in terms of the number of \lya pixels that are included in each estimator per source spectrum, or equivalently how many pixels ``deep" the estimator is along the line of sight.

\section{Simulations and Mock Data}
\subsection{Brief description of DESI}
The DESI survey (\citealt{DESI,DESI2024,DESI_schedule}) is an ongoing large volume spectroscopic redshift survey that will measure over $30$ million objects from $0<z<3.5$ with a 14,000$\sqdeg$ footprint on the sky. Of those objects, $\sim$700,000 will be high redshift quasars ($2.1<z<3.5$) suitable for study of the \lya forest. The resolution of these spectra is $1.8\textrm{ \AA}$ with a \lya pixel size of $0.8\angs$. DESI will represent the largest set of \lya spectra available by a significant margin, the previous being eBOSS (\citealt{eBOSS,eboss20}) with  $~$300,000 \lya spectra.  Table \ref{survey_table} is a brief summary of the relevant parameters of completed and planned \lya surveys. In contrast to previous work ({\citealt{rupert1,ben2,ben1,patrick}}), which mostly considered high angular density \lya spectra, we now consider mock data sets with large footprints and overall number of spectra, but at a lower density. The compromise in angular source density means that recovery of a single lensing potential field with high S/N is not possible. Instead, we will reconstruct lower S/N potentials for the many tiles comprising the total area of the survey on the sky and correlate them with estimates of the lensing potential from foreground galaxy distributions. Future work will involve deriving an estimator for a correlation function between the lensing and the foreground galaxy distribution, but for the present study we will proceed in lensing potential map space to directly compare with previous work.

As described in \cite{ben2,ben1,patrick}, the weak lensing signal from the \lya forest is primarily limited by shot noise in the spectra which propagates uncertainty into the flux contrast pixels. Using the publicly available DESI EDR data (\citealt{DESI_EDR}), we directly computed statistics about the \lya pixel noise. Assuming the final survey will have noise similar to the EDR data, we expect a median pixel noise of $\sigma_\delta\sim0.78$. The pixels are $0.8\angs$, so the noise per $\sqrt{\textrm{\AA}}$ is expected to be $0.87$. The intrinsic \lya fluctuations are expected to have median standard deviation of $\sim0.34$, therefore the expected S/N is fairly low. These noise properties are comparable to those assumed in previous work (\citealt{patrick}) based on the LATIS survey (\citealt{LATIS}), which has pixel noise per $\sqrt{\textrm{\AA}}$ of $\sim0.80$.

In addition to the high redshift \lya quasars, the DESI survey will also measure foreground galaxies: $\sim 10$ million bright galaxy sample galaxies ($0.05<z<0.4$), $\sim4$ million luminous red galaxies ($0.4<z<1.0$), and $\sim17$ million emission line galaxies ($0.6<z<1.6$), for a total sample in excess of $30 $ million galaxies with spectral measurements. The DESI foreground volume also has existing data sets such as the DESI Legacy Survey (\citealt{DESI_legacysurvey}) which contains photometric redshifts for $\sim 50$ million galaxies, should a larger sample of foreground galaxies be necessary for \lya lensing. 

\begin{center}
\begin{table}[h!]\centering%

\begin{tabular}{|c c c c c|}

\hline
Dataset&when&area&Nspectra&density\\
&&($\sqdeg $) &&(deg$^{-2}$)\\

\hline
\multicolumn{5}{|c|}{high density, small area}\\
\hline
CLAMATO&2014-2020&0.5&$\sim 750$&$\sim1{\small,}500$\\
LATIS&2017-2020&1.7&$\sim 3{\small,}800$&$\sim2{\small,}200$\\
Subaru PFS&2025-2029&14&$\sim16{\small,}000$&$\sim1{\small,}150$\\
\hline
\multicolumn{5}{|c|}{low density, large area}\\
\hline

eBOSS&2014-2018&7,500&$\sim 270{\small,}000$&$\sim36$\\
DESI&2021-2026&14,000&$\sim 700{\small,}000$&$\sim50$\\
MSE&2030s-???&80&$\sim160{\small,}000$&$\sim2{\small,}000$\\
\hline

\end{tabular}
\caption{A summary of some relevant parameters for past and future \lya forest surveys (\citealt{CLAMATO,LATIS,subaruPFS,eBOSS,DESI,MSE}) that could be used for weak lensing. We divide the surveys into two general categories: high angular density surveys with smaller footprints and lower density surveys with much larger footprints. DESI, which is the focus of this paper, is a member of the latter category. Though the lensing signal is weaker at larger scales due to weaker correlations in the forest, we are hopeful that the very large number of spectra will compensate.}
\label{survey_table}
\end{table}
\end{center}

\begin{figure}
    \includegraphics[width=\linewidth]{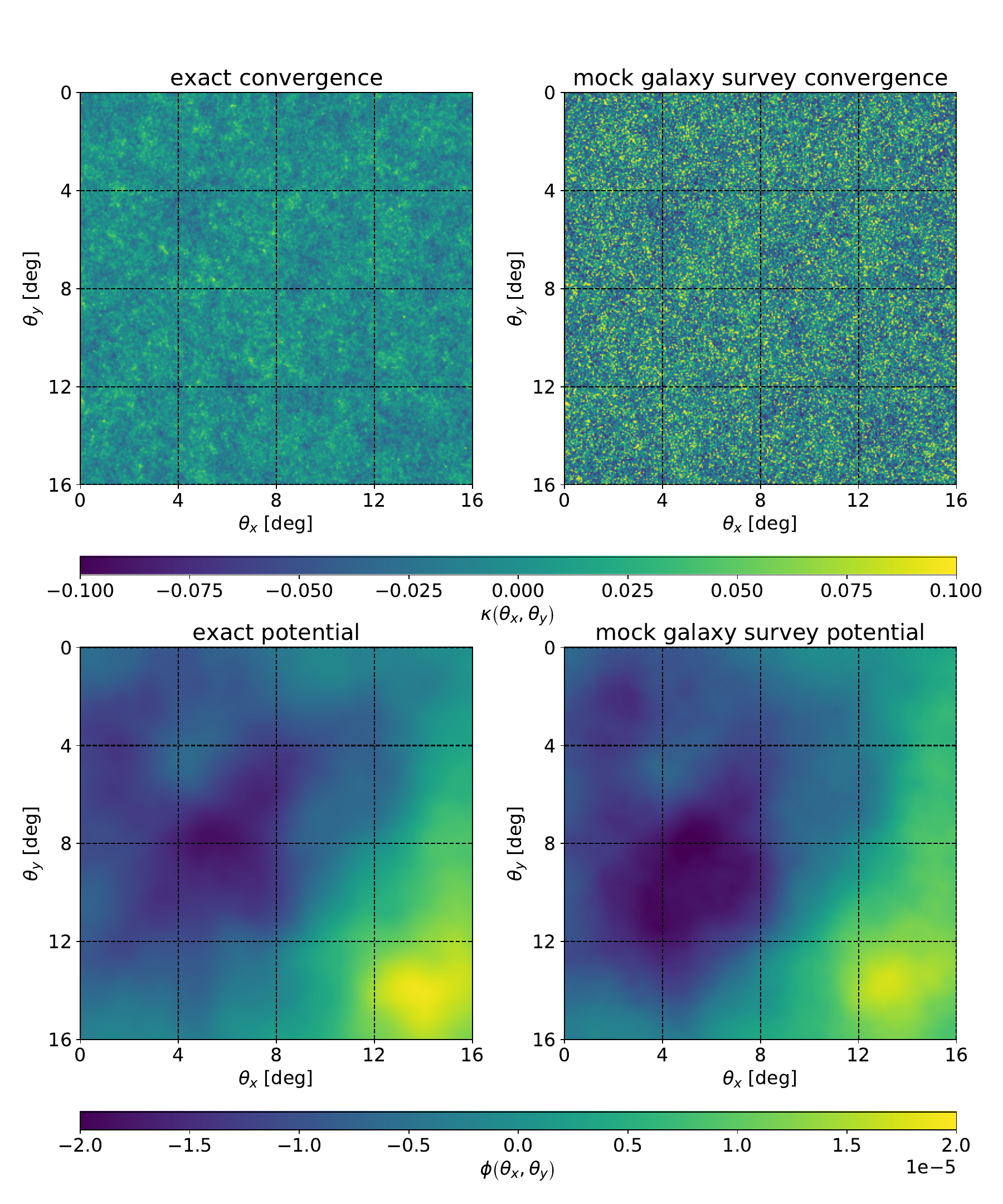}
    \caption{An example simulated foreground field comparing the lensing convergence and potential calculated from the exact matter density contrast field and the contrast field that could be obtained from a 50 million galaxy catalog over the DESI volume. This field represents three independent $1.07 \hmpc$ {\small MP--GADGET} boxes with a resolution of $1024^3$ pixels integrated along the line of sight. The full DESI footprint consists of $50$ such fields, each containing $16$ potential tiles with $256^2$ pixels and area of $16\sqdeg$ that will be reconstructed using the lensing estimator. The Poisson noise and redshift errors add visible noise to the convergence, $\kappa$, but the structure in the potential, $\phi$, remains largely intact. }
    \label{foreground_simulation_fig}
\end{figure}

\subsection{Simulated DESI foreground galaxy distribution}
\label{foreground_simulation_section}
\subsubsection{Foreground density field simulation}
Our first step is to simulate a foreground volume large enough to encompass DESI. Our ultimate goal is to produce realistic lensing potentials as well as mock galaxy survey data similar to what would be available in the real data set. We will then use these data to test how effectively the lensing estimator can produce reconstructions that correlate with the foreground galaxy survey distribution. We produce these mock data through a combination of initial conditions from linear theory and N-body simulation. Although more sophisticated methods of producing mock DESI catalogs have been explored as in \cite{DESI_mock1,DESI_mock2}, for our purposes we preferred a more flexible method in which the number of galaxies in the foreground could be freely varied. 

We set initial conditions using the NGenIC (\citealt{GADGET}) code. We generate initial conditions at $z=2$ for $150$ boxes at a resolution of $1024^3$ with side length $1.07\hgpc$ using $h=0.687$, $\Omega_m=0.2814$, $\Omega_\Lambda=0.7186$, $\sigma_8=0.810, n_s=0.971$ with dark matter only. We start at $z=2$ because it is well below the redshift of the forest and therefore ignores forest self-lensing effects which are not modeled in the estimator. These boxes will be stacked in groups of three, to approximate the length of the line of sight up until the \lya forest begins. The total volume of the simulation is then $\sim184(h^{-1}\textrm{Gpc})^3$ which is approximately equivalent to the 14,000$\sqdeg$ area of DESI over $0<z<2$. This area will ultimately be divided into $800$ square tiles, each with a resolution of $256^2$ pixels and $4\deg$ long on each side.

Next we have to estimate the redshift evolution of the volume along the line of sight. To accomplish this, we time evolve the initial conditions through full N-body dynamics using {\small MP--GADGET} (\citealt{GADGET}). We time evolve a smaller, higher density volume ($267.5\hmpc$ at $1024^3$) from which we obtain the spatial distribution of the particles on the lattice at 8 snapshots, evenly dividing $0<z<2$. By dividing the particle number in each voxel by the mean particle density in the whole volume, we obtain the three-dimensional matter overdensity field, $\delta_\rho$ (as defined in Eq. \ref{delta_rho_equation}), as well as power spectra at each of the eight redshift bins evenly dividing $0<z<2$. The higher resolution is necessary to sufficiently sample each voxel at the final resolution. With a mean of $64$ particles per voxel in the N-body simulation, we have the resolution necessary to recreate the $\delta_\rho$ field at each redshift bin. We then remap the Gaussian distribution of the density field values in each voxel from the linear initial conditions to that of the non-Gaussian distribution of the higher resolution N-body simulations at each of the redshift bins and multiply by the ratios of the power spectra to account for the discrepancy in simulation size. The result is a computationally inexpensive way of approximating full non-linear redshift evolution over the very large volume that DESI will encompass.

Ultimately we obtain $\delta_\rho$ for the entire DESI foreground from which we compute the lensing convergence (Eq. \ref{kappa_equation}) and then the lensing potential (Eq. \ref{potfromkappa_equation}). We approximate the effect of angled lines of sight by shrinking the 2-D slice of the density field being integrated at each integration step along $z$. This introduces the scale mixing that would be observed in real data where sightlines are not parallel. We perform this overall procedure for each of the $50$ simulated volumes, each of which contains $25$ potential tiles that will be reconstructed using our lensing estimator. These $50$ simulated volumes have periodic boundary conditions with themselves, but no boundary conditions are enforced between the volumes. It is worth noting that since the forest itself is lensed using the potential derived from this simulation, the most important source of error is that which is introduced by constructing a mock galaxy survey from the underlying density, rather than the accuracy of the assumptions made to produce the matter density field itself. We will now describe the details of that process.
\subsubsection{Mock galaxy survey}
Simulating the entire volume is useful in that it allows us to produce mock galaxy surveys to which we can compare the results of the estimator. To produce galaxy survey mocks from the matter density field, we first produce a galaxy number density field by Poisson sampling each voxel with expectation given by the matter density field multiplied by the mean galaxy density. The mean galaxy density as a function of redshift is assumed to follow the distribution found in \cite{DESI_nofz}.  We choose a survey size of $1$ million galaxies per box, yielding a total mock survey size of $50$ million galaxies, which is consistent with the $\sim30$ million objects DESI is expected to observe. To be conservative, we still approximate the redshift uncertainty which would arise if the foreground objects only had photometric redshift measurements. We do so by rebinning along the line of sight so that voxels are $\sim 0.05$ in length in redshift space which is the typical uncertainty expected from large volume galaxy surveys (\citealt{euclid_photozerror,SDSS17_photozerror}). However, in the case of a spectral survey like DESI these errors are much smaller.

Finally, we add galaxy bias by assuming the linear bias model described in \cite{galbias}. This model has different bias parameters for redshift $0.4<z<0.6$, $0.6<z<0.8$, and $0.8<z<2.1$. We multiply the $\delta_\rho$ field in each redshift bin by the corresponding bias parameter to obtain a properly biased galaxy survey. From this galaxy survey, we then compute the (biased) galaxy number density field and use that to produce a lensing convergence and lensing potential in the same manner described previously. The impact of the galaxy bias vanishes in the simulated case because we assume the same linear model when we introduce it and then correct for it when computing the lensing potential.

The foreground galaxy distribution potentials will be correlated with the lensing estimator reconstructions, whereas the lensing potentials computed from the exact density contrast field from the initial conditions will be used for lensing the forest. Figure \ref{foreground_simulation_fig} shows the lensing potential that can be recovered from a $50$ million galaxy survey of the DESI foreground compared with the exact result. A substantial amount of Poisson noise can be seen in the convergence map, but when integrating over the surface to compute the potential, most of the prominent features are preserved. We expect that the noise from the \lya forest data will have a much larger impact than the noise introduced in the foreground simulations. The simulation process can be summarized as follows: 
\begin{enumerate}
    \item Generate linear initial conditions, obtaining an exact matter overdensity field, $\delta_\rho$, for the foreground volume of DESI.
    \item Remap the distribution of $\delta_\rho$ obtained from linear initial conditions to the distribution of $\delta_\rho$ which has been evolved through N-body simulation, approximating redshift evolution in $\delta_\rho$ along $z$.
    \item Compute the lensing convergence and potential (Eqs. \ref{potfromkappa_equation}, \ref{kappa_equation} ) from $\delta_\rho$ directly. These will be used to lens the forest. We shrink the size of the map at each integration step to approximate angled lines of sight.
    \item Obtain a mock galaxy survey by converting the redshift-evolved $\delta_\rho$ field into a galaxy number density field. Poisson sample each voxel to add the appropriate counting noise, rebin the voxels along $z$ to approximate photometric redshift noise, and multiply $\delta_\rho$ by the appropriate bias factors.
    \item Compute the lensing convergence and potential from the resulting noisy galaxy survey $\delta_\rho$. This is the lensing potential we would be able to compute from real galaxy data.
\end{enumerate}

\subsection{Gaussian \lya forest}
\label{forest simluation method}
We also need a method for simulating large amounts of \lya forest data. We accomplish this by modeling the forest as a Gaussian random field, as described in \cite{rupert1,ben2,ben1}. The flux contrast pixel correlation function is obtained from the power spectrum fitting function described in \cite{McDonald}, with the redshift of the \lya forest assumed to be $z=2.5$. We perform a Cholesky decomposition of the pixel correlation matrix, $\mathbf{C}$, obtaining 
\begin{align}
   \mathbf{C}=\mathbf{L L}^{T}.
\end{align}
Then we can simulate the pixels directly,
\begin{align}
    \delta_i = \mathbf{L}x_i,
\end{align}
where the $x_i$ are samples from a standard normal distribution. The forest correlation function that is assumed in the quadratic estimator  matches the correlation function used to simulate the data. However, in the case of observational data, it is possible that inaccuracy in the estimate of $\mathbf{C}$ would bias the results of the estimator.

Next, we add Gaussian noise to each $0.8 \angs$ pixel, corresponding to the expected level of noise in the observational data. We compute the median pixel noise of the EDR data (\citealt{DESI_EDR,DESI_EDR_noise}) and excise the noisiest spectra until we obtain a median pixel noise of $\sigma_{\delta}=0.3$.
This level of noise is attained by removing $\sim 25\%$ of the noisiest spectra which maintains a source density well above $50 \persqdeg$. Assuming that the final data are comparable in quality to the EDR, our choice of a $50 \persqdeg$ source density and pixel noise of $\sigma_{\delta}=0.3$ is conservative. Previous work (\citealt{patrick}) has explored the impact that deviations from Gaussianity in the forest have on the efficacy of the estimator. However, given that the scales associated with DESI are significantly larger than those in LATIS (\citealt{LATIS}), with DESI having mean forest sightline separation of $\sim 20 \hmpc$ compared to $\sim 3\hmpc$ in LATIS, we expect the impact of non-Gaussianity to be smaller in the current study. Therefore, for the current analysis we assume a simple Gaussian model for the forest and leave explorations of non-Gaussianity to future work.

\begin{figure*}
    \includegraphics[width=\textwidth]{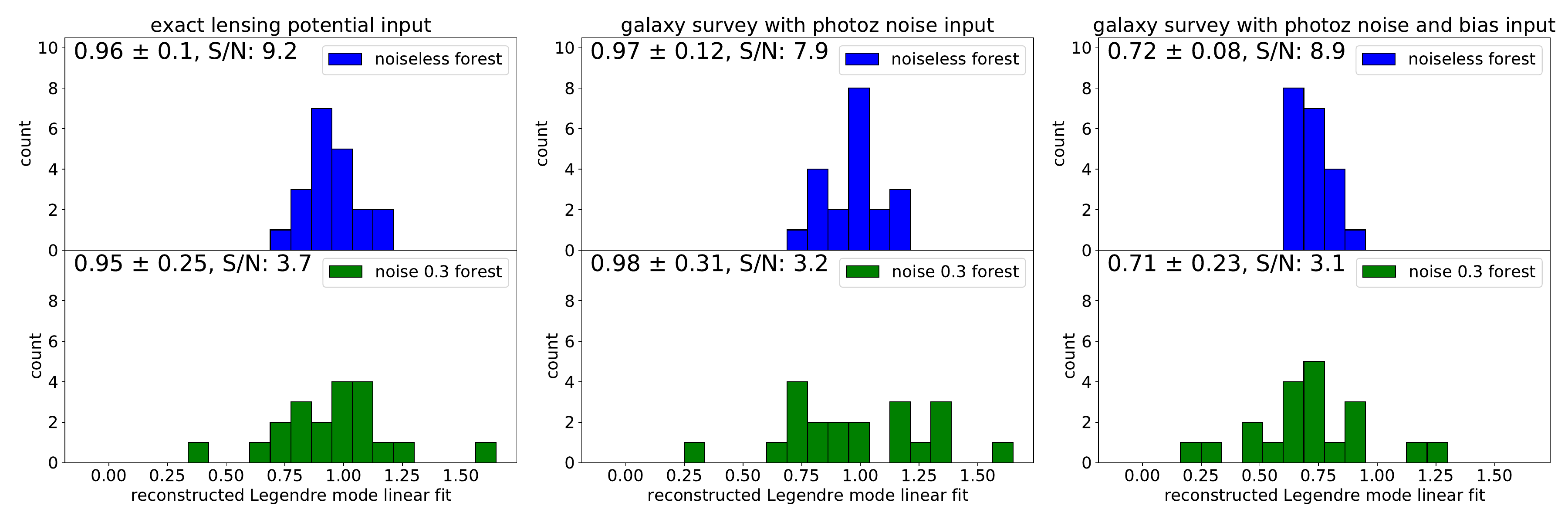}
    \caption{Histograms of the distribution of the linear fit parameters for the Legendre mode amplitudes of the reconstructed lensing potential versus the input lensing potential. We report the mean and standard deviation of each distribution from which we estimate the S/N as the ratio of the mean to the standard deviation. Each column represents a different independent variable for the fits: the first column uses the exact potential that is used to lens the forest, the middle column uses the potential obtained from the mock galaxy survey without galaxy bias (\ref{foreground_simulation_section}), and the right column uses the same mock galaxy survey simulation input but with galaxy bias. The rows represent the amount of noise added to the simulated \lya forest. These results have been corrected for both estimator bias and regression dilution bias using the methods described in Section \ref{bias_section}. The bottom middle panel is our estimate for the full DESI dataset. After lengthening the sightlines to four pixels in each estimator, we expect S/N of $\sim 4$ from the completed DESI data.
    }
    \label{DESI_histograms_plot}
\end{figure*}

\subsection{Quadratic estimator implementation}
 We now describe in detail the overall pipeline for inferring the lensing potential using the quadratic estimator. First, we construct estimators for $100$ randomly generated \lya source geometries over a $\sim16\sqdeg$ field with $800$ sources per field. We use a density of $50$ sources $\persqdeg$ which is the minimum requirement of DESI (\citealt{DESI}).  A total of $800$ such fields are required to fully tile DESI, so each  \lya source geometry is repeated eight times. For computational feasibility,  the estimators are only two pixels deep along $z$.
 In Section \ref{estimatortheory_section} we explain why Monte Carlo averaging of redshift bins is a valid approximation to using full lines of sight. In Section \ref{longersightlines_section} we discuss the impact of including more pixels along $z$ in each estimator. 
 
 The pixels are chosen to be $2.8\angs$ long, the spectral resolution of DESI. However, the DESI \lya pixels themselves are $0.8\angs$ long. Rebinning the DESI pixels to our adopted pixel length reduces the noise in each pixel by a factor of $\sqrt{2.8/.8}$. The noise level of $\sigma_\delta=0.3$ is for the rebinned estimator pixels, longer than the DESI EDR pixels.  
 Next, we simulate $100$ \lya forest spectra using the methods described in Section \ref{forest simluation method} for each source position in the estimator geometry. We ultimately average the results of the $100$ random independent \lya forests, which is equivalent to spectra with $200$ pixels. This is consistent with the sightlines in the DESI EDR, which have median sightline length $\sim 180$ pixels after rebinning (\citealt{DESI_EDR}). We then add uniform Gaussian noise to each pixel by sampling a normal distribution with standard deviation $0.3$, and adding the result to each \lya flux pixel. This is the noise level we find after rebinning to a pixel length of $2.8\angs$ and removing the noisiest $\sim$25\% of spectra from the EDR. After doing so, we still maintain a source density of $\sim 80 \persqdeg$, making our choice of $50 \persqdeg$ in the simulations conservative.
 
 The pixel positions are then lensed according to the exact foreground potential computed using the methods described in Section \ref{foreground_simulation_section}. We use the quadratic estimator to produce estimates of the Legendre mode coefficients for the expansion shown in Eq. \ref{legendre expansion} from each set of $100$ \lya forests. We expand to order five in each dimension which is sufficient to capture the majority of the structure as argued in \cite{ben1}. We ignore the homogeneous and linear modes ($\hat{\phi}_{00},\hat{\phi}_{01},\hat{\phi}_{10}$) which are not detectable from lensing, yielding $22$ mode amplitudes total. Each of these estimates represents one of the $800$ potential tiles in the overall DESI survey. We repeat this process for all $800$ potentials to produce reconstructions for the full DESI volume. Finally, we repeat this entire process $20$ times, effectively simulating $20$ random realizations of the DESI \lya forest data. We do this in order to obtain a distribution of statistics from which we can estimate errors. Specifically, we will compute the distribution of the linear fit parameters of the input potential modes versus the reconstructed potential modes from the lensing estimator.

\begin{figure*}
    \includegraphics[width=\textwidth]{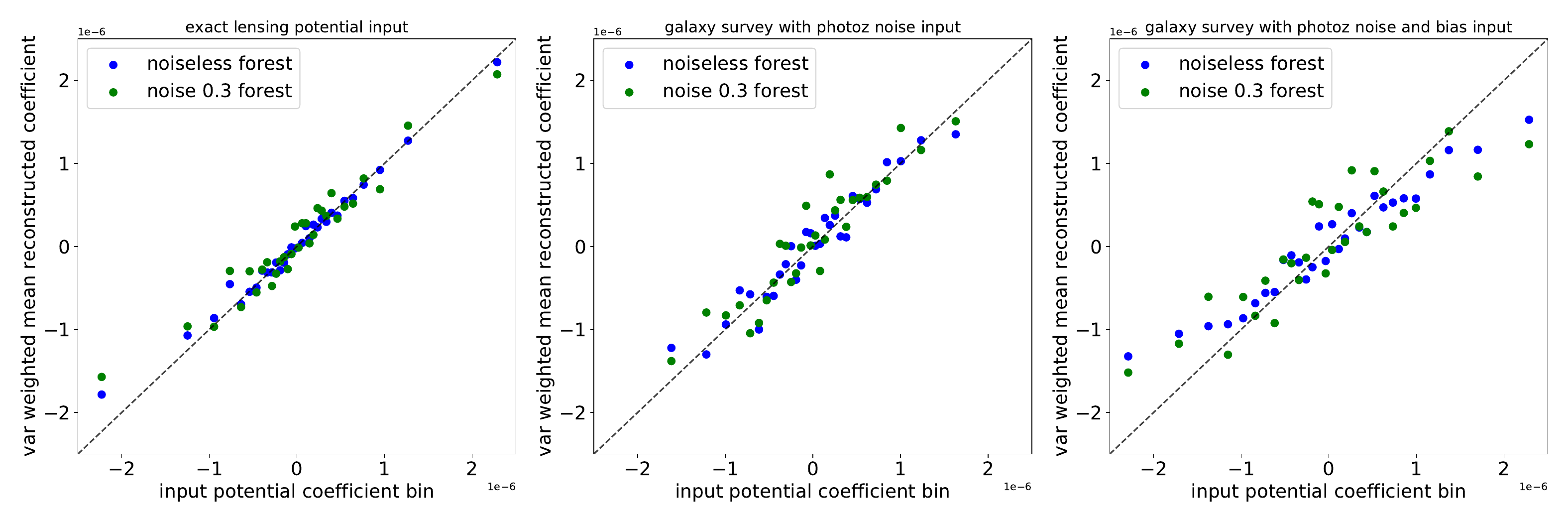}
    \caption{Scatter plots of the reconstructed parameters averaged over $20$ simulated realizations of the DESI \lya forest. We rebin the $22\times800$ reconstructed parameters ($22$ Legendre mode coefficients and $800$ tiles) into $32$ bins with an equal number of data points and average their reconstructions, weighted by the variance of the reconstructions across the $20$ forest realizations. As in Fig. \ref{DESI_histograms_plot}, we show the results with and without noise, as well as for three different independent variables: the exact input lensing potential, the lensing potential as deduced from a mock galaxy survey with redshift uncertainty, and the lensing potential from the mock galaxy survey but with galaxy bias included. The data have been debiased according to the methods described in Section \ref{bias_section}. In the first two panels, we see that the reconstructions follow the $y=x$ line (designated by the dashed black line), indicating the reconstructions are successfully matching the input. In the third panel, the reconstructions are shallower than $y=x$, which is what would be expected when not correcting for the linear galaxy bias.}
    \label{DESI_scatter_plot}
\end{figure*}

\section{Results}
\label{results}

\subsection{Evaluating estimator performance}
\label{statistics_section}
We evaluate the performance of the lensing estimator by computing a S/N ratio for detection. To do so, we perform linear fits of the Legendre mode amplitudes reconstructed by the estimator versus those from the exact lensing potentials as well as the mock galaxy survey potentials. The model that we fit is a one parameter model, with the only parameter being a coefficient multiplying the input mode amplitudes. Therefore, we expect the coefficient fit to be $1$, assuming there is no bias. Each mock DESI survey is comprised of $800$ reconstructed lensing potential fields, each with $22$ Legendre modes, so we fit 17,600 data points in total. We find empirically that the error bars of the fit for the coefficient estimated by the variance weighted least squares fit  underestimate the standard deviation compared to both jackknife and Monte Carlo estimates, likely due to issues with the independence of the mode amplitudes. We find that the Legendre modes are not entirely independent which would lead to inaccuracies in the noise estimate from the linear fit itself. Therefore, we choose to directly simulate $20$ random independent realizations of the DESI data to produce a distribution of parameter fits from which we can estimate the error. We then estimate the S/N of the detection by dividing the mean of the parameter distribution (ideally equal to 1, assuming no bias) by the standard deviation of the fit amplitude distribution. These distributions with their statistics and associated S/N are presented in the histograms in Figure \ref{DESI_histograms_plot}.

The histograms in Figure \ref{DESI_histograms_plot} show the distributions for \lya forests with and without noise, and with three different inputs as the independent variable: the exact lensing potential used to lens the forest in the simulation, the lensing potential computed from the mock galaxy survey without galaxy bias, and the lensing potential computed from the mock galaxy survey with galaxy bias. The bottom middle panel represents the most realistic parameters for DESI. Therefore, based on our simulations, we expect a detection from the real data to have S/N of $\sim 3$, assuming estimators with two pixels per sightline. In Figure \ref{DESI_scatter_plot} we show the scatter associated with these reconstructions, averaged over the 20 realizations of the DESI forest. We rebin the large number of parameters so that the trend can be more easily visualized. We see that the reconstructed parameters scatter about the $y=x$ line closely, indicating that the estimator is successfully reconstructing the input.

In Fig. \ref{fieldreconstructions_figure}, we present a visual representation of our lensing potential reconstructions in real space. Because each individual potential tile is reproduced with low S/N, it is difficult to see the content that is being reconstructed in any given DESI tile. Instead, we average all 800 fields after matching the position of their peaks and do the same for the reconstruction potentials. The peak matching is accomplished by finding the position of the maximum pixel value, and offsetting the tile coordinates so that the maximum pixel value is centered. The potential fields are sufficiently smooth that the position of the maximum pixel value is a good estimate of the position of the central peak. Once offset, we fill the rest of the field in with zeros, the mean potential value, and discard the regions of the field that are now offset beyond the bounds of the $256^2$ pixel image. All 800 DESI tiles are treated this way and then summed and normalized by the number of non-zero pixels that were summed in each pixel of the final image. This method results in an ``average input" that is a central Gaussian peak. When we apply the same method to the reconstructions (offsetting by the peak coordinates of the inputs, not the reconstructions), we can see how well the estimator is doing visually by comparing how well the input Gaussian peak is reproduced in these ``peak-matched" reconstructions. We see that for the noiseless case the estimator tends to reproduce the input peak, whereas once we add noise to the forest the reconstructions are less consistent but still successful on average. We show the reconstructions for each random simulated realization of the DESI forest, so each row is representative of how successful we expect a detection from the real DESI data to be.

Finally, in Figure \ref{legcorrs_fig} we show the Pearson correlation coefficients between the input and reconstructed Legendre mode amplitudes as a function of scale (as defined in Eq. \ref{legendrescale_equation} below). We have $22$ modes total, $9$ of which are degenerate in scale (mode pairs which have $m$ and $n$ exchanged) and $4$ of which are unique ($m=n$), yielding $13$ unique scales. We also mark the corresponding physical length scales as computed at the lensing kernel peak of $z\sim1$ by dividing the length of the tile at $z=1$ by the number of zeros in the Legendre mode being considered. As would be expected, our estimator works better at longer scales than smaller scales.

\subsection{Sources of bias}
\label{bias_section}
\begin{figure*}
    \includegraphics[width=\textwidth]{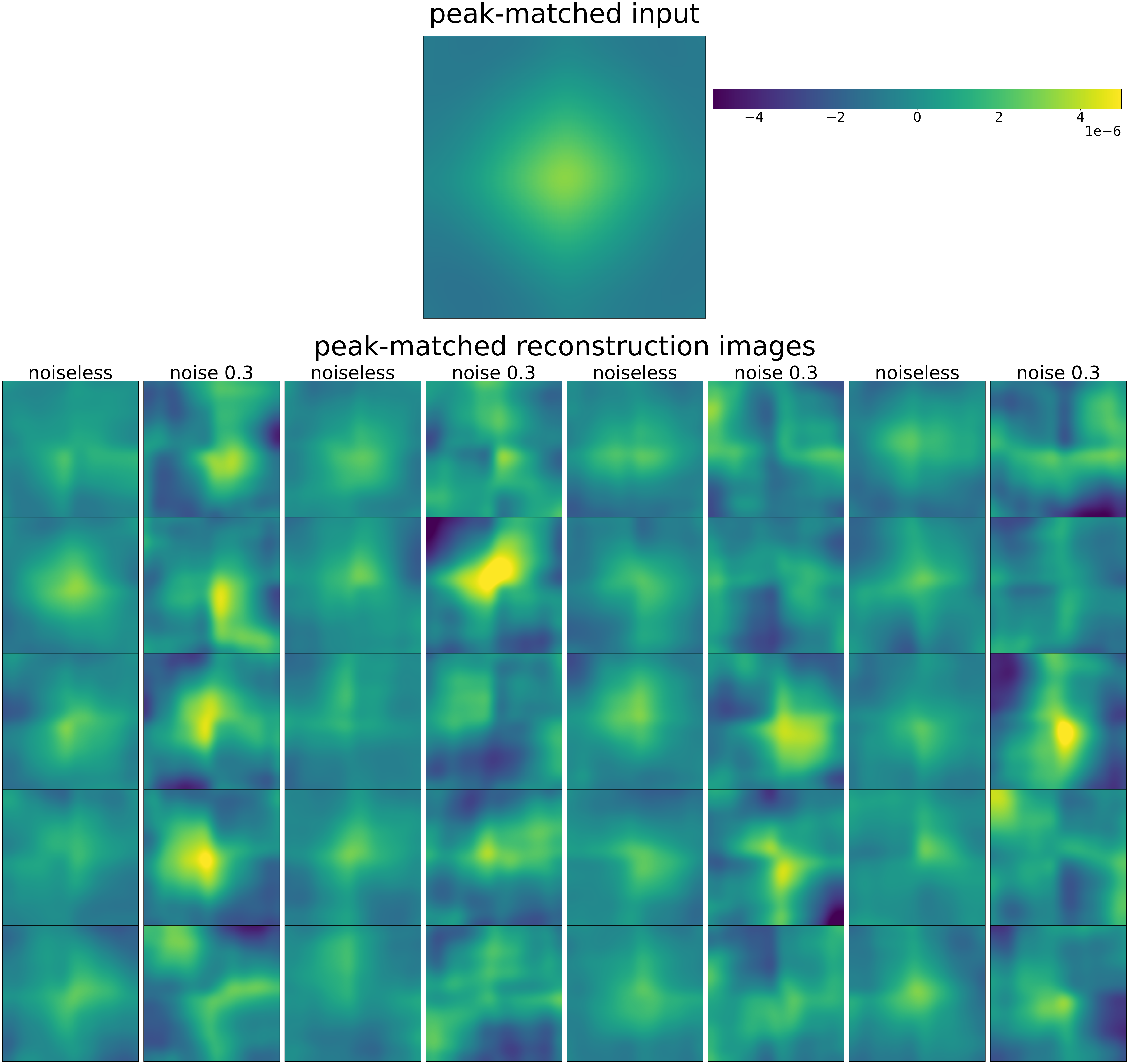}
    \caption{Images of the mean of all 800 potential reconstruction tiles for each forest realization (20 total) of the simulated DESI data with their largest peaks aligned. For each reconstructed field we offset the tiles such that their peaks are centered, average them, and then apply a two-dimensional Gaussian filter with a kernel standard deviation of 10 pixels. The top image is the result of this procedure being applied to the galaxy survey input. The images below are the reconstructions. Each row represents a different random realization of the forest, while the columns represent that realization with and without forest noise.    }
    \label{fieldreconstructions_figure}
\end{figure*}

\begin{figure}
    \includegraphics[width=\linewidth]{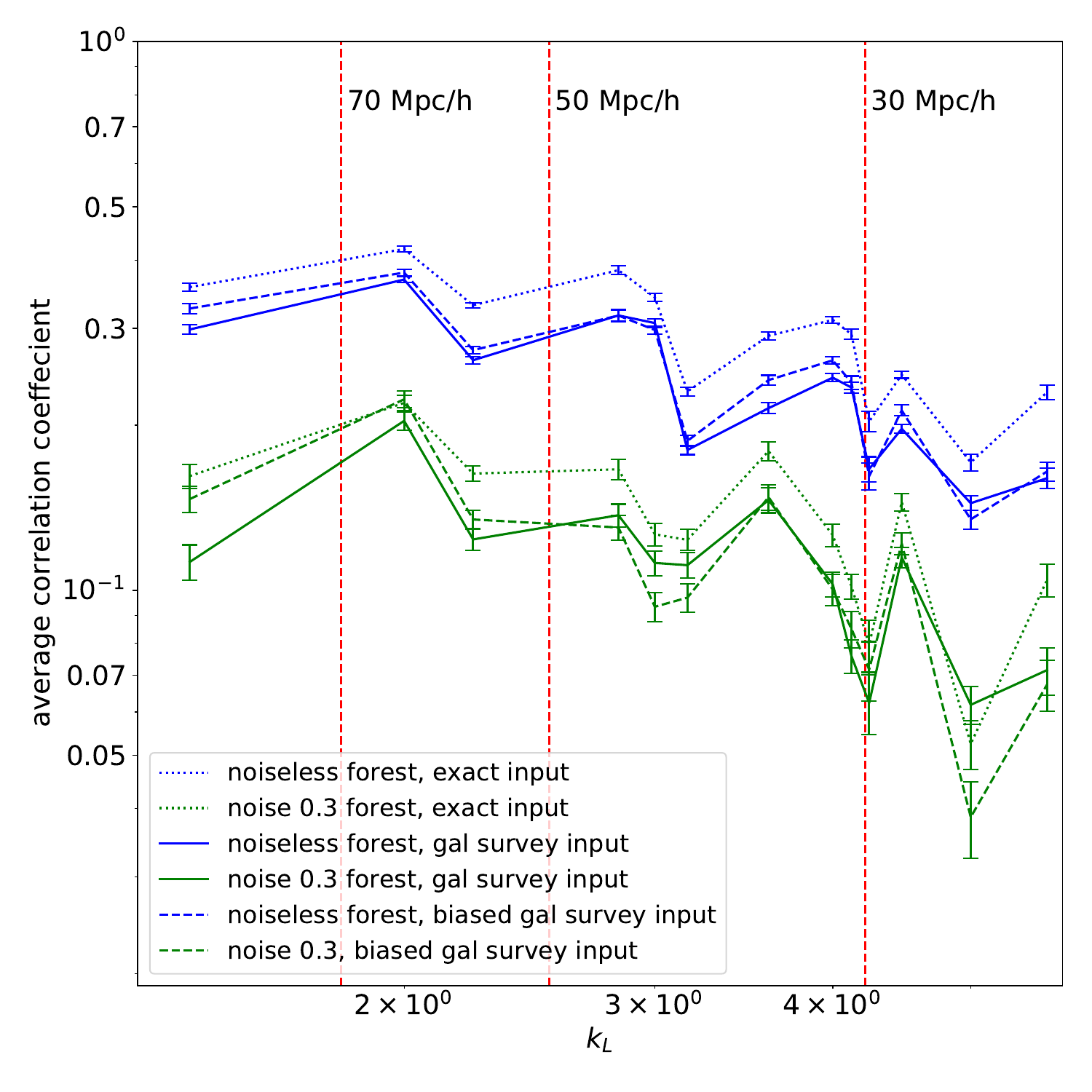}
    \caption{The Pearson correlation coefficients between input and average reconstructed modes as a function of scale, $k_L$, as defined in Eq. \ref{legendrescale_equation}. Associated comoving length scales are indicated by the red dashed lines. These are computed by calculating the size of the field at the lensing kernel peak ($z\sim1$) and then dividing by $k_l$, which gives the number of zeros contained in one field length for those Legendre modes. Error bars on each data point are the jackknife estimates of the standard deviation of the mean of the reconstructions from the 20 forest realizations.}
    \label{legcorrs_fig}
\end{figure}

As described in \cite{patrick}, the estimator appears to be biased empirically. It consistently underestimates the magnitude of the mode coefficients, more dramatically so on smaller scales. Future work will involve quantifying and testing this bias, as well as deriving other lensing estimators to help determine its cause. For the present work we correct for this effect by fitting a line to the multiplicative bias of the estimated parameter magnitudes as a function of scale, where we estimate the scale associated with each mode, $\phi_{mn}P_m\left(x\right)P_n\left(y\right)$, as
\begin{align}
k_{L}=\sqrt{m^2+n^2},
\label{legendrescale_equation}
\end{align}
as suggested in \cite{ben1}. This yields a reasonably good fit (RMSE=$0.14$) of slope $-0.18$ with y-intercept $1.28$, consistent with an unbiased estimate at large scales and an underestimate at smaller scales. We use this fit to correct for the bias in the results presented in Figures \ref{DESI_histograms_plot},\ref{DESI_scatter_plot},\ref{fieldreconstructions_figure}, and \ref{legcorrs_fig}. The impact that this bias correction has on the fits is shown in Table \ref{biascorrections_table}. The S/N estimates change very little because the noise scales along with the signal when applying a multiplicative bias correction. This estimator bias is not to be confused with the galaxy bias, which is a physical bias introduced due to galaxy number density being a biased tracer of the underlying matter distribution. The correction method we have described can be applied to the real dataset as well by running the estimators for the true DESI geometry on simulated forests and lensing potentials.

We also apply a correction (\citealt{regression_dilution}) for the regression dilution bias  introduced when using the galaxy survey potential modes as the independent variable, because they have noise. The slope estimate is divided by a factor of $1-\sigma^2_{\textrm{noise}}/\sigma^2_\textrm{total}$, where $\sigma^2_{\textrm{noise}}$ is the noise introduced to the input modes due to Poisson sampling and redshift uncertainties (estimated from the residuals of the galaxy survey modes compared to the exact input modes) and $\sigma^2_\textrm{total}$ is the total variance in the galaxy survey modes.

\begin{table}[h!]\centering

\begin{tabular}{|c|c c c|}

\hline
&exact&gal survey&gal survey (gal bias)\\
\hline
\multicolumn{4}{|c|}{noiseless}\\
\hline
uncorrected&$0.61\pm0.06$&$0.37\pm0.04$&$0.28\pm0.03$\\
&S/N: 9.8&S/N: 9.4&S/N: 9.0\\
\hline
corrected&$0.96\pm0.1$&$0.97\pm0.12$&$0.72\pm0.08$\\
&S/N: 9.2&S/N: 7.9&S/N: 8.9\\
\hline
\multicolumn{4}{|c|}{noise 0.3}\\
\hline
uncorrected&$0.6\pm0.16$&$0.37\pm0.12$&$0.27\pm0.08$\\
&S/N: 3.7&S/N: 3.2&S/N: 3.3\\
\hline
corrected&$0.95\pm0.25$&$0.98\pm0.31$&$0.71\pm0.23$\\
&S/N: 3.7&S/N: 3.2&S/N: 3.1\\
\hline

\end{tabular}
\caption{A summary of the impact that the bias correction methods described in Section \ref{bias_section} have on the fits of our reconstructions. The bias is essentially eliminated in the first two columns. The bias present in the third column represents the physical galaxy bias. The uncorrected bias in column two appears very severe because it is a combination of both the estimator bias and regression dilution bias. S/N is generally impacted minimally as the method involves multiplying both the signal and the noise by similar amounts. These bias corrections can be implemented in the same manner for the real DESI data by running mock data simulations.}
\label{biascorrections_table}
\end{table}
\subsection{Sources of noise}
Consistent with our expectations from previous work, the noise in the \lya pixels themselves is by far the largest source of uncertainty in the weak lensing measurement. We find that the noise introduced by comparing the results to lensing potentials with Poisson noise and redshift uncertainties is minimal compared to that of the \lya pixels. In Fig. \ref{DESI_histograms_plot}, comparing the bottom left and center bottom panels shows that noise in the \lya spectra reduces S/N by over a factor of 2.5 whereas the noise from the galaxy survey only increases the noise by a factor of $\sim1.2$. We do not expect the noise in the forest pixels to improve significantly in future surveys. However we are optimistic that larger amounts of data and higher densities (as summarized in Table \ref{survey_table}) will be sufficient to mitigate the pixel noise.  

\subsection{Estimators with longer sightlines}
\label{longersightlines_section}
One simplifying assumption we have made throughout our work is that the \lya pixels can be binned into shallow (two pixels deep) redshift bins and treated as independent measurements of the lensing potential without losing too much useful information. This was necessary in order to make the current study computationally feasible. In previous work (\citealt{ben2,ben1,patrick}), where the \lya source densities considered were much larger ($2000\persqdeg$ or higher), the mean lateral separation between pixels was $\sim 3\hmpc$ while the pixel lengths were $\sim 4\hmpc$, meaning that the new pixel pairs introduced by using, for example, $4$ pixels per sightline would have an additional longitudinal separation of $8\hmpc$. As pixel separation increases, their correlations are weaker and therefore contribute less useful information about the lensing. This has been shown empirically in \cite{ben1}. However, with DESI the lateral separations are much larger ($\sim20\hmpc$ on average), so we tested if we could recover more information by making the redshift bins larger. Making the sightlines in each estimator longer is computationally difficult, as the amount of memory required scales as the square of the number of pixels per sightline. However, we were able to carry out a test with four pixels per sightline for a smaller subset of the overall DESI footprint, $200$ potentials rather than $800$. We found that the S/N was improved by a factor of $\sim 1.2$ when comparing the results of the two-pixel estimators to the four-pixel estimators for the same $200$ input potentials. This means that a detection from the real DESI data of $\sim 4$ should be possible using four-pixel estimators with sufficient computational resources. We obtain this estimate by multiplying the improvement we gained in this test ($\sim1.2$) by the result we obtained for the entire DESI survey using two-pixel estimators ($3.2$).  

\subsection{ S/N comparison to previous work}
\label{SNcomparison_section}
In previous work (\citealt{ben1}), the S/N estimate was based on the ratio between the variance of the input and output Legendre mode amplitudes. Briefly, the signal amplitude is computed by simulating 1,000 Gaussian lensing potentials at the desired field size, decomposing them into the Legendre mode basis and then computing the variance of each mode amplitude. The noise amplitude is computed from the variance of the reconstructed modes across the random forest realizations. This approach assumes that the estimator is fully recovering the input signal. In our work, we instead compute S/N based on how successfully each Legendre mode in the input is reconstructed compared to the noise in the estimate by way of a linear fit (see Section \ref{statistics_section}). Our S/N estimates are therefore less optimistic than those reported previously. Figure \ref{SNcomparison_figure} shows a comparison of the two methods for our simulated data. We see that the mean estimate of S/N is a factor of $\sim1.6$ worse when using the linear fit as a S/N estimate.

\begin{figure}
\label{SNcomparison_figure}
    \includegraphics[width=\linewidth]{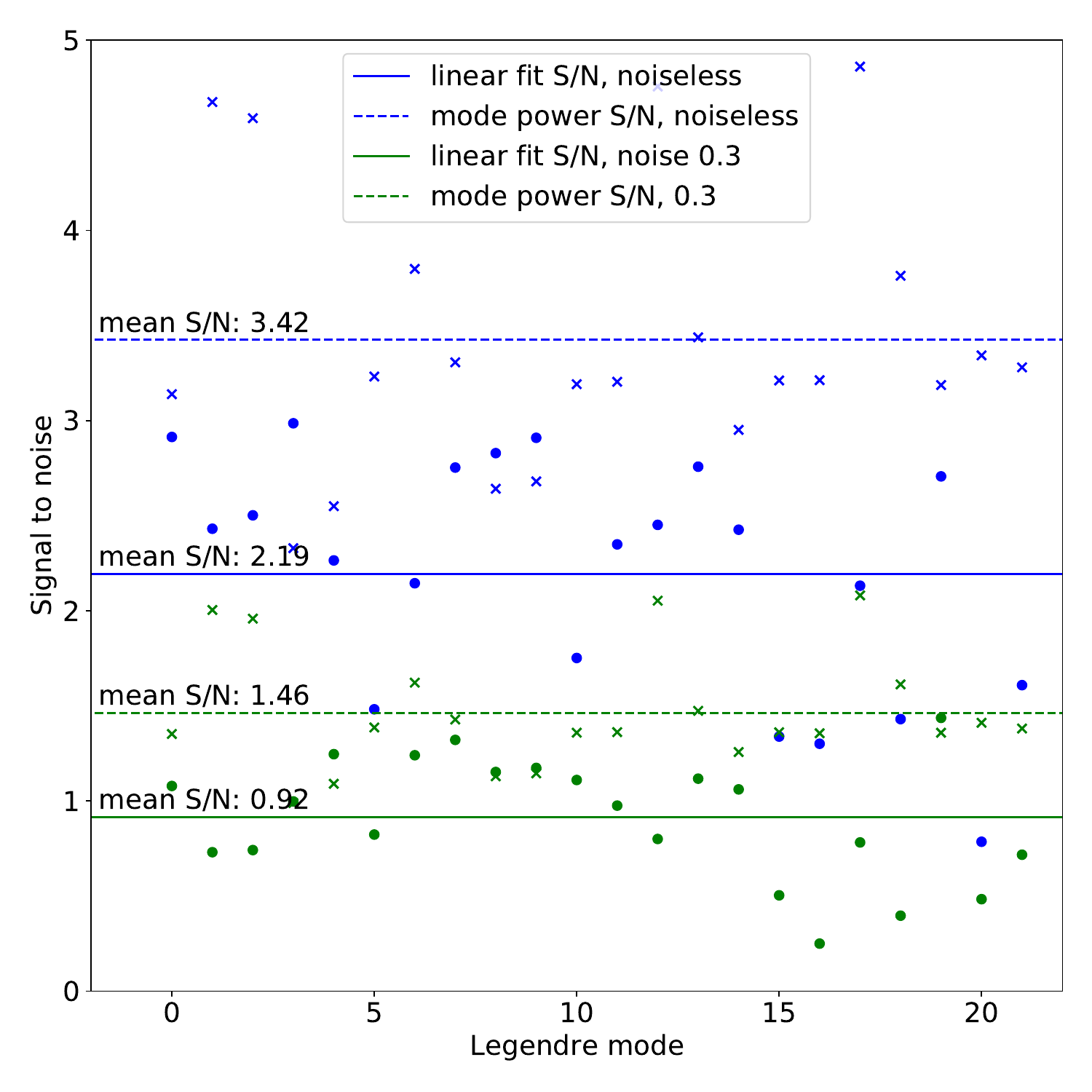}
    \caption{A comparison between the S/N estimates from previous work (\citealt{ben1}) and the current work. The x-axis represents each of the 22 Legendre mode amplitudes reconstructed, going left to right along $\phi_{mn}$ and neglecting the $(0,0)$,$(0,1)$, and $(1,0)$ modes. Modes 0 to 2 correspond to $\phi_{02},\phi_{03},\phi_{04}$, modes 3 to 6 correspond to $\phi_{11},\phi_{12},\phi_{13},\phi_{14}$ and so on. The crosses represent the estimates from the mode power, which is the method used in previous work. The solid dots are the estimates from the linear fit. As described in Section \ref{SNcomparison_section}, the linear fit method which is sensitive to the accuracy of the reconstructions exhibits a lower S/N for each mode.  }
    \label{SNcomparison_figure}
\end{figure}

\begin{table}[h!]\centering
\begin{tabular}{|c c c|}

\hline
density ($\persqdeg$)&S/N DESI footprint&Nspec S/N=10\\
\hline
\multicolumn{3}{|c|}{noiseless}\\
\hline
50&5&$9\times10^{5}$\\
150&15&$7\times10^{5}$\\
200&37&$8\times10^{4}$\\
800&90&$2\times10^{4}$\\
\hline
\multicolumn{3}{|c|}{noise 0.3}\\
\hline
50&2&$8\times10^{6}$\\
150&14&$3\times10^{6}$\\
200&16&$5\times10^{5}$\\
800&27&$1\times10^{5}$\\
\hline

\end{tabular}
\caption{A summary of our estimates for the S/N of a \lya weak lensing detection from hypothetical future surveys with higher source densities. The middle column is the S/N expected if the indicated source density were to be extended over a footprint the size of DESI. The right column is the number of spectra that would be required at the density indicated to measure weak lensing with a S/N of 10. Higher densities require fewer spectra overall because the forest has stronger correlations at those scales. The first row approximates the parameters of DESI to give a point of comparison. }
\label{futuresurveys_table}
\end{table}
\subsection{Future surveys}
In order to make forecasts about the future of \lya lensing, we also perform smaller tests using the geometries of possible future \lya surveys. We simulate just one potential field, at four different source densities (shown in the first column of Table \ref{futuresurveys_table}) and extrapolate the S/N assuming it will scale as the square root of the number of fields. We expect this estimation to be generally accurate, especially for higher density fields. One can compare the results of this method for the 50 source $\persqdeg$ (the DESI source density) to the full results in Figure \ref{DESI_histograms_plot} to find that the S/N estimations are close: 5 and 2 for the noiseless and noise case (respectively) from one field and 8 and 3 from the fully simulated dataset. In Table \ref{futuresurveys_table} we present the S/N we expect from a survey with the indicated source density and a footprint the size of DESI, as well as the number of spectra at a given source density that would be required to obtain a weak lensing detection with S/N of $\sim10$. Comparing to Table \ref{survey_table}, we see that both Subaru PFS and MSE would be well over the requirements for a weak lensing detection with S/N $>10$.

\section{Summary and Discussion}
\label{DiscussionandConclusions}

\subsection{Summary}
We have performed thorough tests of our quadratic estimator in the low source density regime. We simulated realistic foreground lensing potentials and galaxy distributions using approximations of non-linear redshift evolution. With these mock data sets we tested the performance of the estimator over the entire DESI volume using simulated Gaussian random \lya forest data that approximates the expectations of the DESI survey. Comparisons between the foreground galaxy distribution and reconstructions from the estimator show that our estimator is successfully reconstructing the input lensing potential. We estimate the S/N of detection by running our algorithm for 20 realizations of the DESI data and computing the mean and standard deviation of the linear fit of the reconstructed lensing parameters compared to to the foreground galaxy distribution. Doing so shows that for estimators with two-pixel deep sightlines can detect weak lensing in mock DESI data with S/N of $\sim3$ and extending the sightlines to four pixels each improves the result to S/N of $\sim 4$.

\subsection{Discussion}
\label{discussion_section}
We have shown that our quadratic estimator is capable of detecting weak lensing from \lya forest data with low angular source density. More specifically, we have thoroughly evaluated its performance on datasets comparable to what is expected from the DESI survey. While our algorithm is more effective ``per spectrum" at higher source densities, we find that the very large amount of \lya pixels in a survey like DESI is capable of effectively compensating for the angular sparsity. The S/N of $\sim 4$ expected for weak lensing detection from DESI is a significant improvement over what is expected from already existing dense datasets such as CLAMATO or LATIS, where S/N of detection with forest noise is well below unity (\citealt{patrick}). Although the low source density precludes the reconstruction of a single lensing potential field with high S/N, we nonetheless can detect lensing through foreground galaxy distribution correlations. Our statistical forecasts suggest that a detection of S/N$\sim4$ will be possible from the completed DESI data. Our forecasts for future surveys such as MSE or Subaru PFS are even more optimistic with a S/N of well over $10$. Although \lya pixel noise is still relatively high, the large amount of data in these future surveys will be more than enough to overcome it. Although our estimator appears to be biased, relatively simple methods of bias correction that would be applicable to the real data mitigate its impact almost entirely. With this proof of efficacy, \lya lensing represents a promising new method for measuring structure at a unique redshift with different sources of systematic error to existing methods. Overall, we are optimistic about the potential for \lya lensing to serve as a novel, independent probe of late Universe structure growth in the future. The completion of DESI will likely provide a first detection, and subsequent large spectroscopic surveys will increase the S/N drastically.

\subsection*{Data availability}
All data used in this work are available through request to the author.

\subsection*{Acknowledgements}
RACC and PS were supported by NSF NSF AST-1909193, and the NSF AI Institute: Physics of the Future, NSF PHY- 2020295. RBM acknowledges support from MIUR, PRIN 2022 (grant 2022NY2ZRS 001). 


\bibliographystyle{mnras}
\bibliography{ref} 

\end{document}